# Dual-hop transmissions with fixed-gain relays over Generalized-Gamma fading channels

Kostas P. Peppas, Akil Mansour and George S. Tombras

**Abstract**—In this paper, a study on the end-to-end performance of dual-hop wireless communication systems equipped with fixed-gain relays and operating over Generalized-Gamma (GG) fading channels is presented. A novel closed form expression for the moments of the end-to-end signal-to-noise ratio (SNR) is derived. The average bit error probability for coherent and non-coherent modulation schemes as well as the end-to-end outage probability of the considered system are also studied. Extensive numerically evaluated and computer simulations results are presented that verify the accuracy of the proposed mathematical analysis.

✦

## 1 INTRODUCTION

RECENTLY, research efforts have been focused on the investigation of multi-hop wireless communications systems, which seem to extend the coverage without using large power at the transmitter and increase connectivity and capacity in wireless networks [1]–[10]. Multi-hop wireless communications systems are able to provide a potential for broader and more efficient coverage in bent pipe satellites and microwave links, as well as modern ad-hoc, cellular, WLAN, and hybrid wireless networks. In multi-hop networks, intermediate nodes operate as relays between the source and the destination terminal. Generally, there are two main categories of multi-hop wireless communication systems: Non-regenerative and regenerative systems. In the regenerative systems, the relay re-encodes and retransmits the signal towards the destination after demodulating and decoding the received signal from the source. At the destination, the receiver can employ a variety of diversity combining techniques to benefit from the multiple signal replicas available from the relays and the source. Non-regenerative systems use less complex relays that just amplify and and re-transmit the information signal without performing any sort of decoding. Moreover, relays in non-regenerative systems systems can in their turn be classified into two subcategories, namely, channel state information (CSI)-assisted relays and blind relays. Non-regenerative systems with CSI-assisted relays use instantaneous CSI of the first hop to control the gain introduced by the relay. On the other hand, systems with blind relays employ at the relaying nodes amplifiers with fixed gains. Although such systems are not expected to perform as well as systems equipped with CSI-assisted relays, they are characterized by low complexity and ease of deployment.

A versatile fading envelope distribution, which generalizes many of the commonly used models for multi-path and shadow fading, is the generalized gamma (GG) distribution [11]. This fading model is quite general as it includes the Nakagami-$m$ and the Weibull distributions as special cases and the log-normal distribution as a limiting case. Representative past works concerning the performance of dual-hop systems over fading channels can be found in [6], [12], [13]. In [6], the authors have studied the end-to-end performance of dual-hop transmission systems with regenerative and non-regenerative relays, respectively, over Rayleigh fading channels. In [12], the performance of dual-hop wireless communication systems with fixed-gain relays over Nakagami-$m$ fading channels was investigated. Also, in [13], lower bounds of the performance of dual-hop relaying over independent GG fading channels were given. In the view of the appropriateness of the GG distribution for characterizing real-world communication links, it is appealing to inspect the performance of dual-hop systems operating over these channels. However, to the best of the authors' knowledge, results concerning the performance of non-regenerative systems with fixed-gain relays operating over GG fading are not available in the open technical literature.

In this paper a thorough performance analysis of dual-hop wireless communications systems with fixed gain relays is presented. A novel closed form expression for the moments of the end-to-end output SNR is derived. Based on this formula, the outage performance and the average error probability for binary coherent and non-coherent modulation schemes are studied, using the well-known moment-generating function (MGF) approach [14]. The

• K. P. Peppas is with the Laboratory of Mobile Communications, Institute of Informatics and Telecommunications, National Centre for Scientific Research–"Demokritos," Patriarhou Grigoriou and Neapoleos, Agia Paraskevi, 15310, Athens, Greece.
• A. Mansour and G. S. Tombras are with the Department of Electronics, Computers, Telecommunications and Control, Faculty of Physics, University of Athens 15784, Greece.

proposed method for the evaluation of the MGF is based on the Pade approximants theory. Moreover, an alternative integral representation for the outage probability of the considered system is presented. A new closed-form expression is derived for the gain of previously proposed semi-blind relays. These formulae are used in numerical and computer simulations results, to verify the correctness of the presented mathematical analysis. Our results incorporate similar others available in the open technical literature, such as those for Nakagami-$m$ fading channels.

The remainder of the paper is organized as follows: In Section 2, the system and channel model is described in details. In Section 3, closed form expressions for the moments of the end-to-end SNR are presented. In Section 4, the gain of a previously proposed class of semi-blind relays is derived in closed form. In Sections 5 and 6, the error rate and outage performance of the considered system are addressed, respectively. Numerical and computer simulation results are presented in Section 7, while the paper concludes with a summary given in Section 8.

## 2 SYSTEM AND CHANNEL MODEL

We consider a wireless communication system where a source terminal **A** is communicating with a destination terminal **C** through a terminal **B** which acts as a relay. The node **B** amplifies and forwards the received signal to the destination **C** without any sort of decoding. Assuming that the source is transmitting a signal with an average power normalized to unity, the end-to-end SNR is given as [12, Eq. 1]:

$$\gamma_{end} = \frac{(a_1^2/N_{0_1})(a_2^2/N_{0_2})}{(a_2^2/N_{0_2}) + (1/\mathcal{G}^2 N_{0_1})} \quad (1)$$

where $a_i$ is the fading amplitude of the ith hop, $i = 1, 2$, assumed to be GG distributed, $\mathcal{G}$ is the relay gain, and $N_{0_i}$ is the single-sided power spectral density of the additive white Gaussian noise (AWGN) at the $i$-th hop. When blind relays are used, the fixed gain $\mathcal{G}$ established in the connection is $\mathcal{G}^2 = 1/(\mathcal{C}N_{0_1})$, where $\mathcal{C}$ is a constant [6]. Thus (1) becomes:

$$\gamma_{end} = \frac{\gamma_1 \gamma_2}{\mathcal{C} + \gamma_2} \quad (2)$$

where $\gamma_i = a_i^2/N_{0_i}$ is the instantaneous SNR of the ith hop. The probability density function (pdf) of $\gamma_i$ is given by [11]

$$f_{\gamma_i}(\gamma_i) = \frac{\beta_i \gamma_i^{m_i \beta_i/2 - 1}}{2\Gamma(m_i)(\tau_i \overline{\gamma}_i)^{m_i \beta_i/2}} \exp\left[-\left(\frac{\gamma_i}{\tau_i \overline{\gamma}_i}\right)^{\frac{\beta_i}{2}}\right] \quad (3)$$

where $\beta_i > 0$ and $m_i > 1/2$ are parameters related to fading severity, $\overline{\gamma}_i = \mathbb{E}\langle \gamma_i \rangle$ with $\mathbb{E}\langle \cdot \rangle$ denoting expectation, $\tau_i = \Gamma(m_i)/\Gamma(m_i + 2/\beta_i)$ and $\Gamma(x) \triangleq \int_0^\infty e^{-t} t^{x-1} dt$ is the Gamma function. For $\beta_i = 2$, (3) reduces to the Nakagami-$m$ fading distribution whereas for $m = 1$ the Weibull distribution is obtained. Moreover, the cumulative distribution function (cdf) of $\gamma_i$ may be expressed as

$$F_{\gamma_i}(\gamma_i) = 1 - \frac{\Gamma\left(m_i, \left(\frac{\gamma_i}{\tau_i \overline{\gamma}_i}\right)^{\frac{\beta_i}{2}}\right)}{\Gamma(m_i)} \quad (4)$$

where $\Gamma(x, y) \triangleq \int_x^\infty e^{-t} t^{y-1} dt$ is the upper incomplete Gamma function.

## 3 MOMENTS OF THE END-TO-END SNR

In this section, a closed-form expression for the moments of the end-to-end SNR is derived. The $n$-th moment of $\gamma_{end}$ is given by

$$\mathbb{E}\langle \gamma_{end}^n \rangle = \int_0^\infty \int_0^\infty \left(\frac{\gamma_1 \gamma_2}{\mathcal{C} + \gamma_2}\right)^n f_{\gamma_1}(\gamma_1) f_{\gamma_2}(\gamma_2) d\gamma_1 d\gamma_2 \quad (5)$$

Using (3), $\mathbb{E}\langle \gamma_{end}^n \rangle$ can be written as:

$$\mathbb{E}\langle \gamma_{end}^n \rangle = \frac{1}{4} \prod_{i=1}^2 \frac{\beta_i}{\Gamma(m_i)(\tau_i \overline{\gamma}_i)^{\frac{\beta_i}{2}}}$$
$$\times \int_0^\infty \gamma_1^{m_1 \beta_1/2 + n - 1} \exp\left[-\left(\frac{\gamma_1}{\tau_1 \overline{\gamma}_1}\right)^{\frac{\beta_1}{2}}\right] d\gamma_1$$
$$\times \int_0^\infty \left(\frac{\gamma_2}{\mathcal{C} + \gamma_2}\right)^n \gamma_2^{m_2 \beta_2/2 - 1} \exp\left[-\left(\frac{\gamma_2}{\tau_2 \overline{\gamma}_2}\right)^{\frac{\beta_2}{2}}\right] d\gamma_2 \quad (6)$$

The integral with respect to $\gamma_1$, $\mathcal{I}_1$, can be evaluated by applying the change of variables $\left(\frac{\gamma_1}{\tau_1 \overline{\gamma}_1}\right)^{\frac{\beta_1}{2}} = t$ and using the definition of the $\Gamma$ function as

$$\mathcal{I}_1 = 2 \frac{(\tau_i \overline{\gamma}_i)^{m_1 \beta_1/2 + n}}{\beta_1} \Gamma\left(m_1 + \frac{2n}{\beta_1}\right) \quad (7)$$

The integral with respect to $\gamma_2$, $\mathcal{I}_2$, can be evaluated by expressing the exponential and the fraction in terms of Meijer-G functions i.e. $\exp(-x) = G_{0,1}^{1,0}[x \mid \frac{-}{0}]$ [15, Eq. 8.4.3.2] and $(1+x)^{-\rho} = \frac{1}{\Gamma(\rho)} G_{1,1}^{1,1}\left[x \mid \frac{1-\rho}{0}\right]$ [15, Eq. 8.4.2.5] and with the application of [15, Eq. 2.24.1.1] as:

$$\mathcal{I}_2 = \frac{\sqrt{k_2} l_2^{n-1} \mathcal{C}^{m_2 l_2}}{\Gamma(n)(2\pi)^{l + \frac{k_2 - 3}{2}}}$$
$$\times G_{l_2, k_2 + l_2}^{k_2 + l_2, l_2}\left[\frac{(\tau_2 \overline{\gamma}_2)^{-l_2} \mathcal{C}_2^l}{k_2^{k_2}} \bigg| \begin{matrix} \Delta(l, 1 - m_2 l_2 - n) \\ \Delta(k_2, 0), \Delta(l_2, -m_2 l_2) \end{matrix}\right] \quad (8)$$

where $k_2$ and $l_2$ are the minimum integers that satisfy $\beta_2 = 2l_2/k_2$ and $\Delta(x, a) = \{\frac{a}{x}, \frac{a+1}{x}, \ldots \frac{a+x-1}{x}\}$. It is noted that the Meijer-G function is a standard built-in function available in the most popular software-based mathematical packages such as Maple or Mathematica. Finally, the moments of the end-to-end SNR of the considered system are given in closed-form as

$$\mathbb{E}\langle \gamma_{end}^n \rangle = \frac{\Gamma\left(m_1 + \frac{2n}{\beta_1}\right)(\tau_1 \overline{\gamma}_1)^n l_2^n \mathcal{C}^{m_2 l_2}}{\sqrt{k_2}\Gamma(m_1)\Gamma(m_2)(\tau_2 \overline{\gamma}_2)^{\frac{m_2 l_2}{k_2}} \Gamma(n)(2\pi)^{l_2 + \frac{k_2-3}{2}}} \\ \times G_{l_2,k_2+l_2}^{k_2+l_2,l_2}\left[\frac{(\tau_2 \overline{\gamma}_2)^{-l_2} \mathcal{C}_2^l}{k_2^{k_2}} \bigg| \begin{matrix}\Delta(l_2, 1-m_2 l_2 - n) \\ \Delta(k_2, 0), \Delta(l_2, -m_2 l_2)\end{matrix}\right] \quad (9)$$

By substituting $n = 1$ to (9) a closed-form expression for the average end-to-end SNR can be obtained. For the special case of Nakagami-$m$ fading channels ($\beta_1 = \beta_2 = 2$), it can be observed that (9) is reduced to a previously known result [12, Eq. 9].

## 4 A Class of "Semi-Blind" Relays in generalized-gamma fading channels

In this section, a new expression for the gain of a -previously published- class of "semi-blind" relays is presented in closed form for GG fading channels. In [6], the authors proposed a specific class of "semi-blind" relays which consume the same average power with the corresponding CSI-based relays. The proposed fixed gain relay, benefits from the knowledge of the first hop average fading power. In such a scenario, the fixed gain is considered equal to the average of CSI assisted gain, namely

$$\mathcal{G}^2 = \mathbb{E}\left\langle \frac{1}{a_1^2 + N_{0_1}^2} \right\rangle \quad (10)$$

For GG fading, by performing the required statistical average in (10) and following a process similar to the one for the evaluation of $\mathcal{I}_2$, $\mathcal{G}$ can be expressed in closed form as

$$\mathcal{G}^2 = \frac{l_1}{N_{0_1}(2\pi)^{l_1 + \frac{k_1-3}{2}}\sqrt{k_1}\Gamma(m_1)(\tau_1 \overline{\gamma}_1)^{m_1 l_1/k_1}} \\ \times G_{l_1,k_1+l_1}^{k_1+l_1,l_1}\left[\frac{(\tau_1 \overline{\gamma}_1)^{-l_1}}{k_1^{k_1}}\bigg|\begin{matrix}\Delta(l_1,1-m_1 l_1)\\ \Delta(k_1,0),\Delta(l,1-m_1 l_1)\end{matrix}\right] \quad (11)$$

where $k_1$ and $l_1$ are the minimum integers that satisfy $\beta_1 = 2l_1/k_1$. Finally, the parameter $\mathcal{C}$ can be obtained as $\mathcal{C} = 1/(\mathcal{G}^2 N_{0_1})$. For Nakagami-$m$ fading channels ($l_1 = k_1 = 1, \tau_1 = 1/m_1$), by making use of the identity $G_{1,2}^{2,1}\left[x\big|_{0,1-b}^{1-a}\right] = \Gamma(a)\Gamma(a-b+1)\Psi(a,b,x)$ [15, Eq. 8.4.46.1] where $\Psi(\cdot,\cdot,x)$ denotes the Tricomi hypergeometric function [15, Eq. (7.2.2.7)] and $\Psi(a,a,z) = e^z \Gamma(1-a,z)$ [15, Eq. (7.11.4.4)], we observe that $\mathcal{G}^2$ reduces to a previously known result [12, Eq. (12)].

## 5 Padé Approximants and Average Bit Error Probability

In this section we address the error performance of the considered dual-hop system for different coherent and non-coherent binary modulation schemes. The ABEP of various digital modulation schemes over fading channels can be evaluated by using the well-known MGF based approach [14]. In this case, however, it is difficult to derive a closed form expression for the MGF of the output SNR, $\mathcal{M}_{\gamma_{end}}(s)$. Instead, it is more convenient to use the Padé approximants method [16], a simple and efficient method to accurately approximate the MGF and in sequel to evaluate the ABEP. The main advantage of this method is that due to the form of the produced approximation, the ABEP can be calculated directly using simple expressions for the non-coherent Binary Frequency Shift Keying (BFSK) and Binary Differential Phase Shift Keying (BDPSK) modulation schemes, while for M-ary Quadrature Amplitude Modulation (QAM) and M-ary Phase Shift Keying (PSK), single integrals with finite limits and integrands composed of elementary functions can be readily evaluated by numerical integration.

A Padé approximant to the MGF is a rational function of a specified order $B$ for the denominator and $A$ for the nominator, whose power series expansion agrees with the $(A+B)$-order power expansion of the MGF, namely

$$\mathcal{M}_{\gamma_{end}}(s) \simeq R_{[A/B]}(s) = \frac{\sum_{i=0}^{A} c_i s^i}{1 + \sum_{i=0}^{B} b_i s^i} \simeq \sum_{n=0}^{A+B} \frac{\mathbb{E}\langle \gamma_{end}^n \rangle s^n}{n!} \quad (12)$$

where $b_i$ and $c_i$ are real numbers. In order to obtain an accurate approximation of the MGF, we assume sub-diagonal Padé approximants ($B = A+1$) [16]. The coefficients $b_i$ and $c_i$ may be numerically evaluated using any of the most popular commercial software mathematical packages such as Maple or Mathematica.

Using (12), the ABEP of digital modulations for several signaling constellations may be efficiently evaluated. For example, the ABEP of BDPSK can be readily obtained from (12) as $\overline{P}_{be} = 0.5 \mathcal{M}_{\gamma_{end}}(-1)$. Also, the ABEP for coherent binary signals is given by [14]

$$\overline{P}_{be} = \frac{1}{\pi}\int_0^{\pi/2} \mathcal{M}_{\gamma_{end}}\left(-\frac{\psi}{\sin^2\theta}\right)d\theta \quad (13)$$

where $\psi = 1$ for coherent binary phase shift keying (BPSK), $\psi = 1/2$ for coherent BFSK and $\psi = 0.715$ for coherent BFSK with minimum correlation.

## 6 End-To-End Outage Probability (OP)

In this section the end-to-end Outage Probability (OP) of the considered system is addressed. The OP is defined as the probability that the instantaneous output SNR, $\gamma_{end}$, falls below a specified threshold $\gamma_{th}$. Two methods for the evaluation of the end-to-end OP are presented: The first method is based of the Padé approximants theory where as the second one derives an easy-to-evaluate integral representation of the OP.

## 6.1 Evaluation of the end-to-end OP using Padé approximants

The outage probability can be extracted from $\mathcal{M}_{\gamma_{end}}(s)$ based on the following Laplace transformation

$$P_{\text{out}}(\gamma_{\text{th}}) = \mathbb{L}^{-1}\left\{\frac{\mathcal{M}_{\gamma_{end}}(s)}{s}; s; t\right\}\bigg|_{t=\gamma_{\text{th}}} \quad (14)$$

where $\mathbb{L}^{-1}\{\cdot, s; t;\}$ denotes inverse Laplace transform. Using (12) and the residue inversion formula [17], the OP can be obtained as

$$P_{out}(\gamma_{th}) = 1 - \sum_{i=1}^{B}\frac{\lambda_i}{p_i}e^{p_i\gamma_{th}} \quad (15)$$

where $p_i$ are the poles of the of the Padé approximants to the MGF, which must have negative real part, and $\lambda_i$ are the residues.

## 6.2 An integral representation of the end-to-end OP

Using (2) the end-to-end OP may be obtained as

$$\begin{aligned}P_{\text{out}}(\gamma_{\text{th}}) &= \Pr(\gamma_{end} \leq \gamma_{\text{th}}) = \\ &\int_0^\infty \Pr\left(\frac{\gamma_1\gamma_2}{\mathcal{C}+\gamma_2} \leq \gamma_{\text{th}}\bigg|\gamma_2\right)f_{\gamma_2}(\gamma_2)d\gamma_2 \\ &= \int_0^\infty \Pr\left(\gamma_1 \leq \frac{(\mathcal{C}+\gamma_2)\gamma_{\text{th}}}{\gamma_2}\bigg|\gamma_2\right)f_{\gamma_2}(\gamma_2)d\gamma_2 \\ &= 1 - \int_0^\infty\left[1 - F_{\gamma_1}\left(\frac{(\mathcal{C}+\gamma_2)\gamma_{\text{th}}}{\gamma_2}\right)\right]f_{\gamma_2}(\gamma_2)d\gamma_2 \\ &= 1 - \frac{\beta_2}{2\Gamma(m_1)\Gamma(m_2)(\tau_2\overline{\gamma}_2)^{m_2\beta_2/2}} \\ &\quad \times \int_0^\infty \Gamma\left(m_1,\left(\frac{\mathcal{C}\gamma_{\text{th}}+\gamma_{\text{th}}\gamma_2}{\tau_1\overline{\gamma}_1\gamma_2}\right)^{\frac{\beta_1}{2}}\right)\gamma_2^{m_2\beta_2/2-1} \\ &\quad \times \exp\left[-\left(\frac{\gamma_2}{\tau_2\overline{\gamma}_2}\right)^{\frac{\beta_2}{2}}\right]d\gamma_2\end{aligned} \quad (16)$$

where $\Pr(\cdot)$ denotes the probability operator. A closed form expression for the previously defined integral is very difficult, if not impossible to be obtained. However, after performing the change of variables $\gamma_2 = \tau_2\overline{\gamma}_2 w^{\frac{2}{\beta_2}}$, the OP may be expressed after some algebraic manipulations as

$$\begin{aligned}P_{\text{out}}(\gamma_{\text{th}}) = &1 - \frac{1}{\Gamma(m_1)\Gamma(m_2)}\int_0^\infty w^{m_2-1}e^{-w} \\ &\times \Gamma\left[m_1,\left(\frac{\mathcal{C}\gamma_{\text{th}}+\gamma_{\text{th}}\tau_2\overline{\gamma}_2 w^{\frac{2}{\beta_2}}}{w^{\frac{2}{\beta_2}}\prod_{j=1}^2 \tau_j\overline{\gamma}_j}\right)^{\frac{\beta_1}{2}}\right]dw\end{aligned} \quad (17)$$

It can be observed that this integral can be accurately and efficiently evaluated by using the Gauss-Laguerre quadrature rule [18]. Thus, the end-to-end OP may be

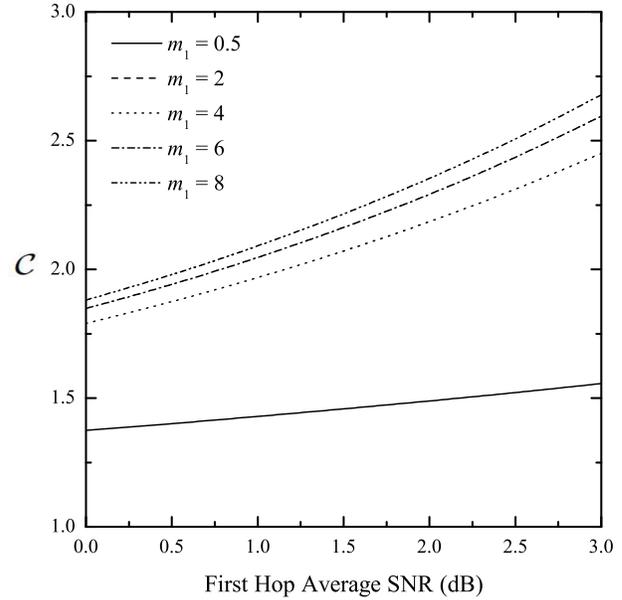

Fig. 1. Parameter $\mathcal{C}$ versus $\overline{\gamma}_1$ for $\beta_1 = 4/3$ and for several values of $m_1$.

obtained as

$$\begin{aligned}P_{\text{out}}(\gamma_{\text{th}}) &\simeq 1 - \frac{1}{\Gamma(m_1)\Gamma(m_2)} \\ &\times \sum_{i=1}^N W_i x_i^{m_2-1}\Gamma\left[m_1,\left(\frac{\mathcal{C}\gamma_{\text{th}}+\gamma_{\text{th}}\tau_2\overline{\gamma}_2 x_i^{\frac{2}{\beta_2}}}{x_i^{\frac{2}{\beta_2}}\prod_{j=1}^2 \tau_j\overline{\gamma}_j}\right)^{\frac{\beta_1}{2}}\right]\end{aligned} \quad (18)$$

where $x_i$ are the roots of the $N$-th order Laguerre polynomial $L_N(x)$ and $W_i$ are the corresponding weights given by

$$W_i = \frac{x_i}{(N+1)^2[L_{N+1}(x_i)]^2} \quad (19)$$

## 7 NUMERICAL AND COMPUTER SIMULATION RESULTS

In this section, various performance evaluation results obtained by numerical and simulations techniques that illustrate the formulations derived herein are presented.

In Fig. 1, the parameter $\mathcal{C}$ as a function of $\overline{\gamma}_1$ for several values of $m_1$ and $\beta_1 = 4/3$ is depicted and as expected, $\mathcal{C}$ increases as $m_1$ increases. In Fig. 2, the average end-to-end SNR as a function of $\overline{\gamma}_1$ for $\beta_1 = \beta_2 = 3$ and $m_1 = m_2 = 2$ is depicted. In the same plot, the impact of power imbalance between the two hops on the considered metric is also illustrated. As expected [6], when $\overline{\gamma}_2 > \overline{\gamma}_1$, it is beneficial and, otherwise, it is detrimental. In Fig. 3, the ABEP for BDPSK and BPSK is illustrated as a function of $\overline{\gamma}_1$ for balanced ($\overline{\gamma}_2 = \overline{\gamma}_1$) and unbalanced ($\overline{\gamma}_2 = 2\overline{\gamma}_1$) hops assuming $\beta_1 = \beta_2 = 3$ and $m_1 = m_2 = 2$. As it is evident, ABEP decreases as $\overline{\gamma}_1$ increases. Moreover, in Figs. 4 and 5 the impact of the fading parameters $\beta$ and $m$ on the ABEP is illustrated. More specifically, in Fig. 4 the ABEP of BDPSK and BPSK is illustrated for $\overline{\gamma}_2 = 2\overline{\gamma}_1$, assuming $\beta_1 = \beta_2 = 2.5$ and $m_1 = m_2 = m$, as a function of $\overline{\gamma}_1$ and for $m = 1.5, 3$ and $3.5$. One can observe that ABEP improves as $m$ and/or $\overline{\gamma}_1$ increases. In Fig. 5, ABEP results for BDPSK and BPSK are presented for $\beta_1 = \beta_2 = \beta$, $m_1 = m_2 = 2$, $\beta = 1, 2.5, 3.5$ and as it is obvious the ABEP improves as $\beta$ and/or $\overline{\gamma}_1$ increases.

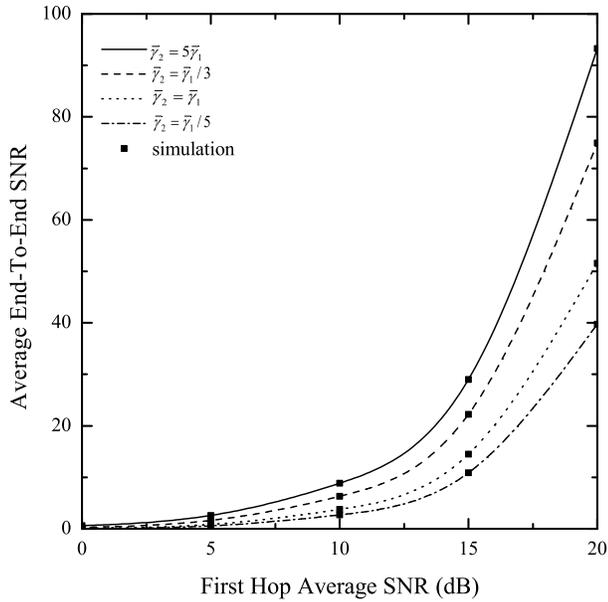

Fig. 2. Average end-to-end SNR versus $\overline{\gamma}_1$ for $\beta_1 = \beta_2 = 3$ and $m_1 = m_2 = 2$.

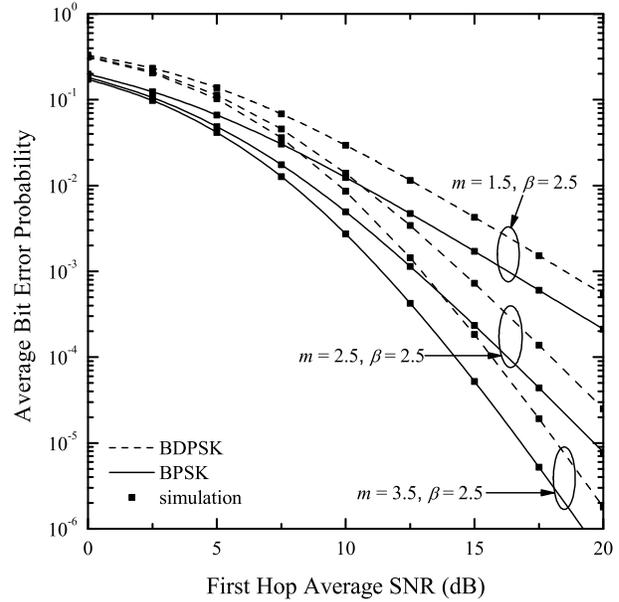

Fig. 4. Average Bit Error Probability versus $\overline{\gamma}_1$ for unbalanced ($\overline{\gamma}_2 = 2\overline{\gamma}_1$) hops, $\beta_1 = \beta_2 = 2.5$, $m_1 = m_2 = m$ and for various values of $m$.

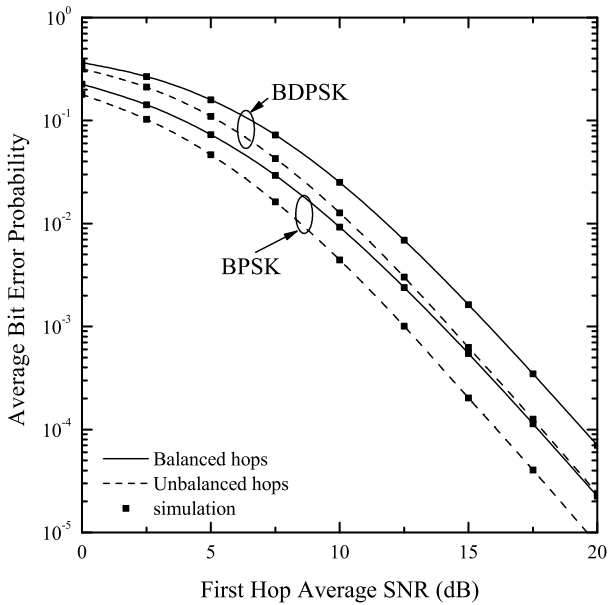

Fig. 3. Average Bit Error Probability versus $\overline{\gamma}_1$ for balanced ($\overline{\gamma}_2 = \overline{\gamma}_1$) and unbalanced ($\overline{\gamma}_2 = 2\overline{\gamma}_1$) hops ($\beta_1 = \beta_2 = 3$ and $m_1 = m_2 = 2$).

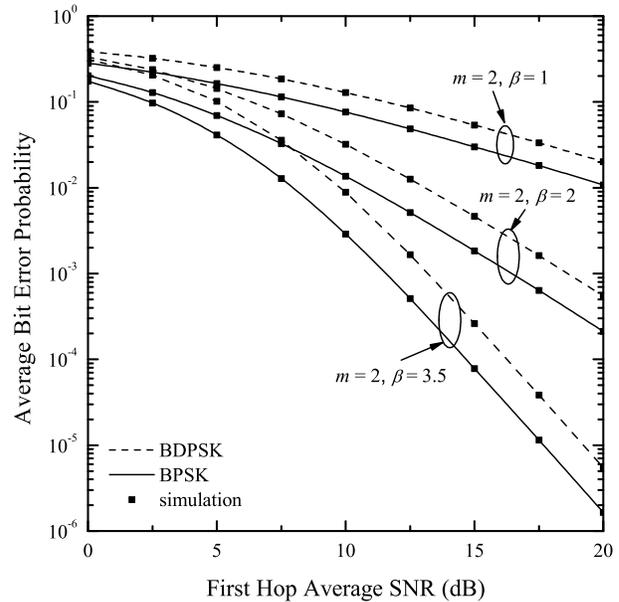

Fig. 5. Average Bit Error Probability versus $\overline{\gamma}_1$ for unbalanced ($\overline{\gamma}_2 = 2\overline{\gamma}_1$) hops, $\beta_1 = \beta_2 = \beta$, $m_1 = m_2 = 2$ and for various values of $\beta$.

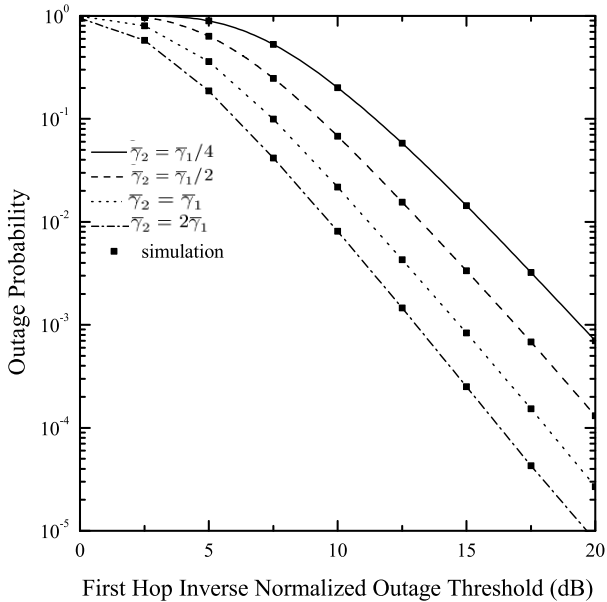

Fig. 6. Outage Probability versus $\overline{\gamma}_1/\overline{\gamma}_{th}$ ($\beta_1 = \beta_2 = 3$ and $m_1 = m_2 = 2$).

In Fig. 6, the OP of the considered system is illustrated as a function of the first hop inverse normalized outage threshold $\overline{\gamma}_1/\overline{\gamma}_{th}$ when $\beta_1 = \beta_2 = 3$ and $m_1 = m_2 = 2$. The impact of power imbalance between the two hops on the OP is also illustrated. As far as the power imbalance is concerned, one can verify similar findings to that mentioned in Fig. 2. Finally, for all the considered test cases, our theoretical analysis is substantiated by means of monte-carlo simulations and as it can be observed, the simulations are in perfect agreement with the analytically obtained results.

## 8 CONCLUSION

In this paper, the end-to-end performance of dual-hop wireless communication systems with fixed-gain relays operating over GG fading channels was evaluated. A novel closed-form expression for the moments of the output SNR was derived. Moreover, the average error and the outage performance of the considered system were studied using the MGF approach and the Padé approximants method. An alternative integral representation for the outage probability was also derived. This expression can be acurrately and efficiently evaluated by means of the Gauss-Laguerrre quadrature rule. Various numerical and computer simulations results were presented that demonstrated the proposed mathematical analysis.

**Kostas P. Peppas** was born in Athens in 1975. He obtained his diploma in Electrical and Computer Engineering from the National technical University of Athens in 1997 and the Ph.D. degree in telecommunications from the same department in 2004. His current research interests include wireless communications, smart antennas, digital signal processing and system level analysis and design. He is a member of IEEE and the National Technical Chamber of Greece.

**Akil Mansour** was born in Deir Ezzor, Syria, in 1969. He received the B.Sc. degree in physics from Aleppo University of Syria, in 1995, the M.Sc. degree in the field of Electronics and Telecommunication systems at the department of Electronics and Telecommunications from National and Kapodistrian University of Greece, Athens, in 2001. At the present time he is doing Ph.D. in the field of Wireless Mobile Communication.

**George S. Tombras** was born in Athens, Greece, in 1956. He received the B.Sc. degree in physics from Aristotelian University of Thessaloniki, Greece, the M.Sc. degree in electronics from University of Southampton, UK, and the Ph.D. degree from Aristotelian University of Thessaloniki, in 1979, 1981, and 1988, respectively. He is currently an Associate Professor of Electronics. His research interests include mobile communications, analog and digital circuits and systems, as well as instrumentation, measurements, and audio engineering. Professor Tombras authored or coauthored more than 70 journal and conference papers and many technical reports.